# Quantifying the hierarchical scales of scientists' mobility

Yurui Huang[1†], Langtian Ma[1†], Chaolin Tian[1], Xunyi Jiang[1], Roberta Sinatra[2,3,4], Yifang Ma[1*]

[1]Department of Statistics and Data Science, Southern University of Science and Technology, Shenzhen 518055, Guangdong, China

[2]Copenhagen Center for Social Data Science (SODAS), University of Copenhagen, Denmark

[3]Networks, Data, and Society (NERDS) group, IT University of Copenhagen, Denmark

[4]ISI Foundation, Turin, Italy

*Email: mayf@sustech.edu.cn

[†]These authors contributed equally to this work.

**Abstract**

Human behaviors, including scientific activities, are shaped by the hierarchical divisions of geography. As a result, researchers' mobility patterns vary across regions, influencing several aspects of the scientific community. These aspects encompass career trajectories, knowledge transfer, international collaborations, talent circulation, innovation diffusion, resource distribution, and policy development. However, our understanding of the relationship between the hierarchical regional scale and scientific movements is limited. This study aims to understand the subtle role of the geographical scales on scientists' mobility patterns across cities, countries, and continents. To this end, we analyzed 2.03 million scientists from 1960 to 2021, spanning institutions, cities, countries, and continents. We built a model based on hierarchical regions with different administrative levels and assessed the tendency for mobility from one region to another and the attractiveness of each region. Our findings reveal distinct nested hierarchies of regional scales and the dynamic of scientists' relocation patterns. This study sheds light on the complex dynamics of scientists' mobility and offers insights into how geographical scale and administrative divisions influence career decisions.

Scientists' mobility plays an important role in advancing scientific communication, knowledge diffusion, and the advancement of science and technology across cities, countries, and continents [1, 2]. Understanding the mobility patterns of researchers is crucial for various reasons, including scientific career development [3-8], geographical knowledge propagation [9, 10], and fostering both international and domestic collaborations [11, 12]. Historically, human activities like scientific mobility were shaped by geographical factors, leading to a hierarchical arrangement of mobility patterns among cities, countries, and continents [13]. Economic disparities, cultural differences, and varied policies across nations contribute to unequal and biased movements of scientists and knowledge exchange worldwide. Challenges like visa restrictions, geographical distance, language barriers, and cultural variances exacerbate these imbalances [9].

Consequently, scientists' propensity of movement is often restricted by the geographical boundaries. For instance, given the similar geographical distance, the likelihood of domestic movements is higher than that of cross-border movements. Therefore, achieving a thorough understanding of the scientific mobility patterns at various scales, and comprehending the influence of geographical hierarchical scales on such mobility, is crucial. This understanding bears significant implications for scientists' career development, cross-regional knowledge circulation, and policy development, among other areas.

Established studies have found that the mobility of scientists is shaped by a variety of factors and constraints. Traditionally, mobility was deterred by geographical distance. Studies have found that greater geographical distance decreases the propensity of mobility between two regions, with the regional characteristics like language, culture, and economic factors also playing significant roles [14, 15]. Given the complexity of scientists' mobility, researchers have made efforts to investigate the latent patterns. For instance, Gargiulo and Carletti (2014) used data from the American Physical Society to analyze researcher trajectories and applying network theory and complex system analysis, and found that institutions with linguistic similarities between countries tend to exchange researchers more frequently than those without such similarities [16]. Vaccario, Verginer and Schweitzer (2020) explored the temporal correlations in their mobility patterns at various levels of aggregation, highlighting the importance of institutional path dependence in scientific careers and the potential implications for mobility programs [17].

Additionally, various modeling frameworks for general human mobility patterns prove useful in understanding scientists' mobility. The gravity model, known for its straightforward approach, serves as a foundational framework to simplify the complex mobility patterns and reveals the geographical distance as an obstruction in mobility [18, 19]. The radiation model [20] captures the density of populations in each location along the path from the departure location to the destination. The container model [13] considers nested geographical scales in the mobility network. Agent-based models delve into nuanced decision-making processes and emerging behaviors [21, 22]. Furthermore, there are also developed frameworks that use natural language processing (NLP) techniques, such as word embedding of mobility networks and lower-dimensional vector spaces to represent high-dimensional mobility data [23-25]. These models provide foundations for developing scientist mobility models [16, 26-29].

However, there is still a lack in the analysis of hierarchical regional scales when it comes to understanding scientists' mobility [30-32]. By integrating established frameworks for human mobility analysis, we aim to deepen our understanding of scientific mobility. Our approach involves developing a model that captures the intricate, multiscale patterns of scientists' geographical mobilities throughout their careers. In most countries, administrative divisions based on historical and geographical factors are in the form of a hierarchical structure ranging from cities to provinces or states. These divisions, whether between cities, provinces, states, countries, or continents, significantly influence the movement of scientists. While scientists are often drawn to cities with robust economies and superior educational resources [33], they also face challenges such as cultural and lifestyle differences, adaptation of family, and research environment changes after movements.

The large-scale database, OpenAlex, provides detailed records of publications, journals, authors, and institutions [34, 35], offer us a unique opportunity to quantitatively analyze the mobility patterns of scientists. Drawing inspiration from recent advances in human mobility studies, we formulate a model that incorporates geographical divisions and natural administrative regions. In our model, we classify the institutions, cities, countries, and continents involved in scientists' mobility according to their respective administrative levels, treating them as regions of scientific activities. This model enables us to estimate the attractiveness of each region and the propensity for mobility between regions within these nested administrative divisions, with strong implications for science system development.

## Results

**Data processing**
We extracted records of scientists' location mobility records from the OpenAlex database, which includes papers published from 1960 to 2022 and encompasses details on authors, institutions, and publication dates, covering a total of 75,560 academic institutions [34]. We tracked mobility by identifying instances where a researcher publishes a paper at one institution and subsequently publishes a paper at another, considering this as a movement from the initial institution to the latter (details provided in the Methods section). Given that a small proportion of institutions are responsible for the majority of scientific publications, we began by ranking institutions based on their associated publication count. We selected the top 8,366 institutions that contributed to 95% of the papers (See Fig. S1 in SI). We then filtered the 200 major cities as city-level regions. This was achieved by first determining the top 20 countries based on their number of institutions, and then within each of these countries, selecting the top 10 cities by institutional count (see Fig. 1A). We segmented the data into 5-year periods to track the evolving trends (see Methods).

**Evidence of hierarchical scales in scientists' mobility**
We show that the nested hierarchical scales among regions impede free movements. Indeed, in a setting without geographical constraints, we would expect the flux of mobility between different regions to be proportional to the number of papers and the number of citations within each region, as shown by the configuration model [36, 37], which randomizes flux between nodes while preserving the in-degree and out-degree of each node. We performed the degree-preserving configuration modeling at the city, country, and continent levels. This configurational model represents the flows unconstrained by regions or geographical border, but only driven by the differences in production of scientists and willingness to move. We calculated the ratio between the real flux from region $a$ to region $b$, $e_{a,b}$, and the expected flux from the configuration model, $e'_{a,b}$. If the ratio $e_{a,b}/e'_{a,b}$ is less than 1, the flux from region $a$ to region $b$ is larger than expected (over-representative); otherwise, it is under-representative.

Our analysis revealed that mobility flows across boundaries were less frequent compared to those within boundaries. As illustrated in Fig. 1(B–D), the kernel density estimations of the ratios between the real flux and the expectation by the configuration model show large differences within and across boundaries at the city (Fig. 1B), country (Fig. 1C), and continent (Fig. 1D) levels. Scientists prefer to relocate between institutions within the same cities, countries, or continents, rather than traversing between them. This pattern underscores the role of spatial scale in shaping scientists' mobility patterns.

These pronounced differences motivated us to construct a hierarchical framework for modeling scientists' mobility. As shown in Fig. 1E, scientists' mobility activities are encapsulated within various regional scales and are located globally and classified with nested scales such as cities, countries, and continents. The movement from one institution to another may occur at the same level (e.g., move from Harvard to MIT within Boston) or across multiple levels (e.g., move from George Washington University to the University of London, which transfers all three levels of city, country, and continent). Our model captures the different propensities within and across levels and assesses the attractiveness of the destination regions to those relocating movers.

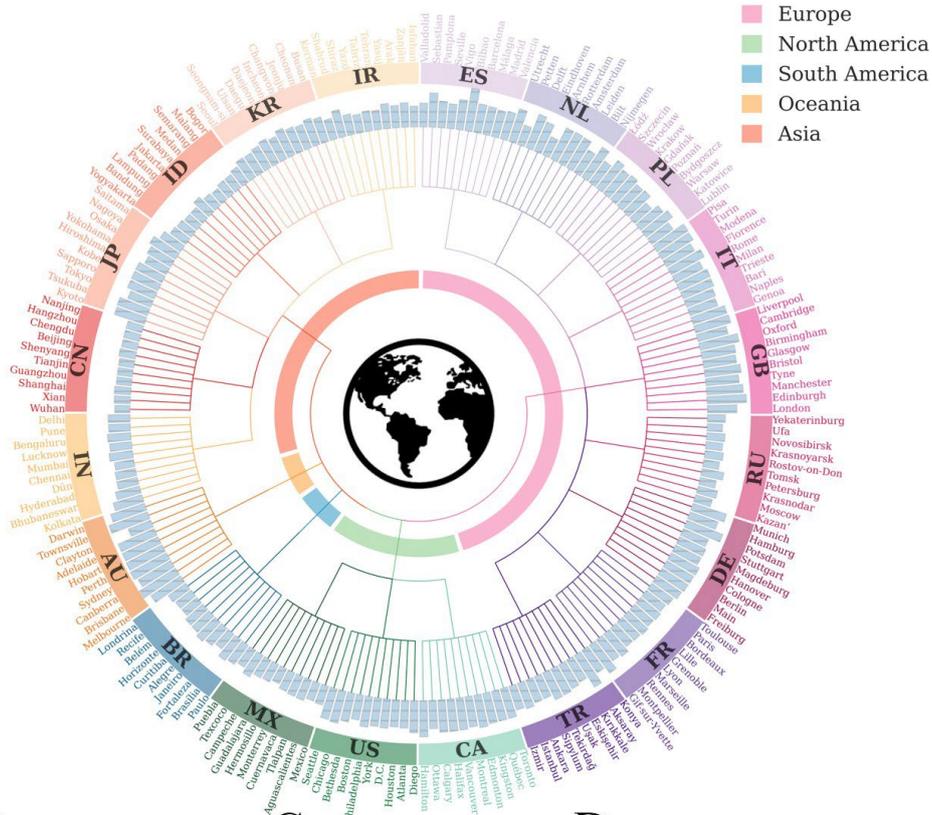
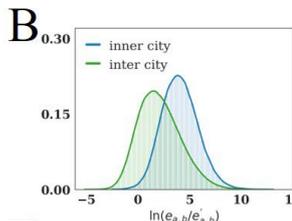
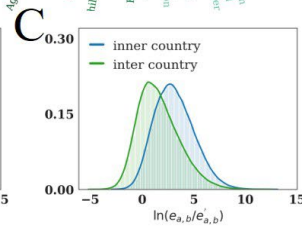
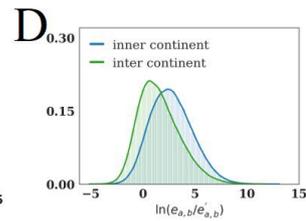
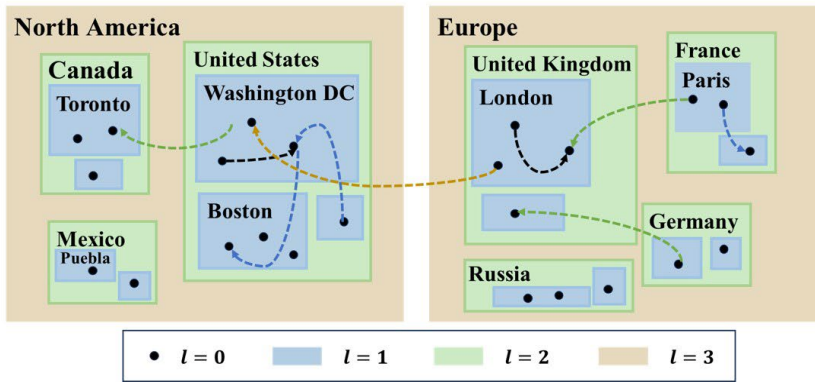

**Figure 1. The scales of scientists' mobility. A.** The hierarchical geographical structure of major cities involved in the mobility of scientific labor worldwide. The dendrogram presents the hierarchical structures from continents, countries, and then to cities. The histogram for each city indicates the average number of publications per author in the city. **B–D.** The kernel density function (KDE) of the ratio, $e_{a,b}/e'_{a,b}$, between real mobility flow and the configuration model expectation, for pairs of institutions within and across cities (B), countries (C), and continents (D). **E.** The hierarchical scale

framework: Researchers transition between institutions (represented by black dots) within the nested regions. Rectangles with different colors represent regions at different hierarchical levels. Dashed arrows in color illustrate potential transitions across different level distances. The probability of transition between two institutions is determined by a combination of two factors: the level distance of the transition and the attractiveness of the destination.

**The SMART Model**
In our Scientific Mobility and Administrative Regions (SMART) model, each scientist moves in a hierarchical physical space with four levels: institutions, cities, countries, and continents, ordered from smallest to largest (Fig. 1E). At any level, space is partitioned into geographical regions, therefore, a lower-level region is fully included within a single-parent region (e.g., each institution with level 0 is part of a city with level 1). Hence, each geographical location $k$ can be identified as a sequence of nested regions $T(k) = (k_0, k_1, k_2, k_3)$, where region $k_l$ is included in $k_{l+1}$, $l$ is the level.

The level distance, $d(j, k)$ between institutions $j$ and $k$ is defined as the highest index (start from 0) at which $j$ and $k$ differ. For example, let $j$ be Beijing Normal University, and $k$ be Tokyo University, then
$$T(j) = (\text{Beijing Normal University, Beijing, China, Asia}),$$
and
$$T(k) = (\text{Tokyo University, Tokyo, Japan, Asia}).$$
Both tuples differ in the first three elements, therefore, the level distance $d(j, k) = 2$.

Based on this setting, the model was characterized by the following parameters [13]:
1. The attractiveness of the nested regions $k$ is denoted as $a(k)$. Attractiveness $a(k_l)$ represents the probability of visiting region $k_l$ among all regions nested within $k_{l+1}$. Consequently, $\sum_{i \in k_{l+1}} a(i) = 1$.
2. The likelihood of traveling at distance $d(j, k)$ starting from country $c$ is $p(d(j, k), c)$, where $c = j_2$ is the country of $j$.

According to our model, the probability of a scientist moving from city $j$ to city $k$ can be written as

$$p(j \to k) = p(d(j,k), c) \prod_{l=0}^{d(j,k)} a(k_j). \tag{1}$$

Here, $p(d(j, k), c)$ denotes the probability of traveling at level distance $d(j, k)$ given the country of $j$ being $c$, that is, the country in which institution $j$ is located. $a(k_l)$ represents the attractiveness of the region $k_l$ (Fig. 1E).

By virtue of the maximum likelihood estimation, we fitted the model and obtained the attractiveness and transition probabilities as defined by Equation (1) (see details in the Methods section). The estimation was performed for each of the 5-year periods from 1960 to 2021.

**Dynamics of region attractiveness**

The model enables us to estimate the attractiveness of each region for every 5-year snapshot, which reflects the region's potential to attract scientists. This attractiveness could be influenced by factors such as status of the scientific environment, prestige in scientific entrepreneurship, or the availability of funding and policy incentives that attract researchers. In Fig. 2, we show the temporal trends of attractiveness of continents, leading countries, and major cities.

Fig. 2A shows that Europe was more attractive than North America and Asia before the 2000s, afterwards, Europe's attractiveness declined, while at the same time Asia's attractiveness steadily ascended. This trend continued until 2015–2021, during which Asia and Europe exhibited similar attractiveness, which was slightly higher than that of North America. Additionally, the attractiveness of Oceania and South America remained relatively stable. Finally, after narrowing our focus to the top 20 countries with the highest number of institutions worldwide, African nations do not appear on the list due to a relatively lower institutional density compared to other continents. This limited availability of data for African countries restricts our analysis of the continent's attractiveness in the context of scientists' mobility.

In Fig. 2(B–D), we show the dynamics of the attractiveness of countries on each continent. In Europe (Fig. 2B), most countries' attractiveness has gradually stabilized over time, with the United Kingdom (GB) losing its early dominance in the 1960s. Spain (ES) has gained a notable position in Europe's talent migration network. In Asia in Fig. 2C, China's (CN) attractiveness surged from the 1980s, and ultimately become Asia's the most appealing nation in Asia, this is due to the rapid economic growth which amplified its talent demand, and China has launched series of policies targeting overseas talents, for instance the Thousand Talent Program. Simultaneously, we witnessed a decline in attractiveness in Japan (JP) and India (IN) in recent decades. Figure 2D shows the trends for the United States (US), Canada (CA), and Mexico (MX), their attractiveness is relatively stable, and the US maintains the highest attractiveness for scientific movers.

Regarding to intra-national attractiveness dynamics, for instance, in China, Beijing, as a longstanding educational hub, has witnessed a gradual shift of talent attraction towards other Chinese cities. In the US, the appeal of cities to talent has gradually transitioned away from the situation where Washington dominates, with Boston emerging as a new focal point for scientific movements. In the United Kingdom, the attractiveness of different cities has been relatively balanced since the 1960s (see Fig. S3A).

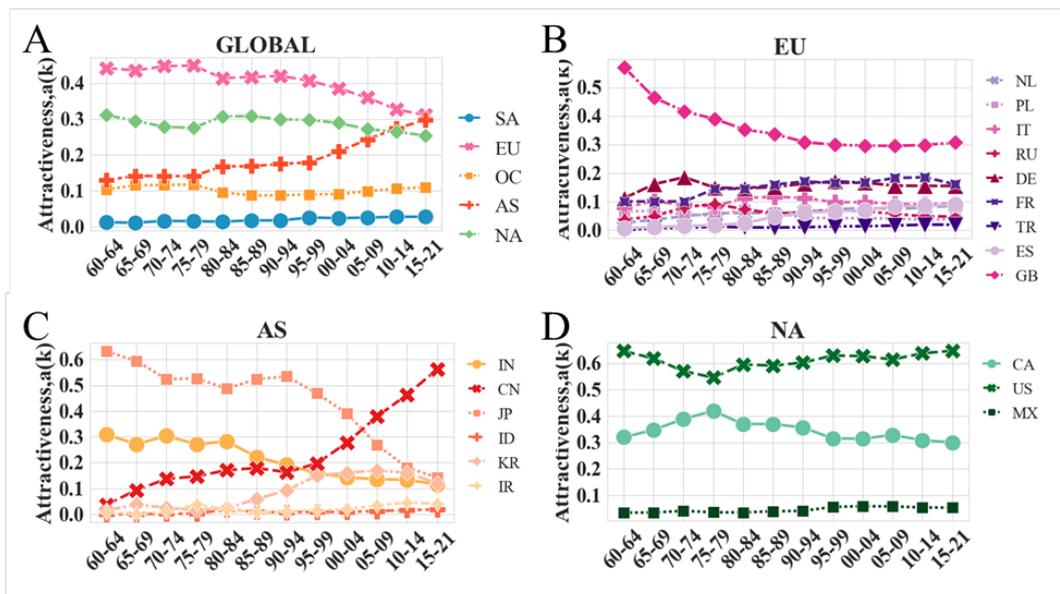

**Figure 2. The estimated attractiveness of regions at different administrative levels.**
**A.** Attractiveness, as measured by the model, at continent level calculated every 5 years from 1960 to 2021. Different colors, markers shape, and line type identify different continents, including South America (SA), Europe (EU), Oceania (OC), Asia (AS), and North America (NA). **B–D.** Attractiveness of different countries in EU (B), AS (C), and NA (D).

**Mobility propensity across different scales**
In Fig. 3A, we illustrate the likelihood of mobility based on their level distances and find that long-distance mobility is declining. Over the past decade, there has been a noticeable trend of scientists opting for employment opportunities within the same country (level distance of 1), and often within the same city (level distance of 0), rather than engaging in cross-border (level distance of 2) or intercontinental moves (level distance of 3). From 2015 to 2021, about 70% of scientists' mobility was concentrated within national borders. This tendency, when considering our utilization of the concept of 'home countries', is in line with the previously observed scale law of mobility, in which scientists prefer to migrate at levels below a certain threshold.

Our model facilitates the estimation of scientists' propensity to move across different distances, starting from their home country. In Fig. 3B, for different home countries where the scientists are located, we plotted the propensity of the scientists' mobility at different distances. A closer examination of the data in Fig. 3B reveals distinct mobility patterns among scientists from different regions. Notably, researchers in the United States tend to have a strong inclination towards international mobility (level distance is greater than or equal to two), often seeking opportunities in countries across the globe. In contrast, scientists in European nations, such as Russia and France, predominantly opt for relocation within their own national borders, indicating a preference for intra-country movements.

Asian scientists, including those from China, generally prefer to stay within their home countries for career advancements. However, Chinese scientists exhibit a unique pattern: they tend to either remain within China or move across continental boundaries, primarily to North America or Europe, rather than to other Asian destinations. These regional disparities in scientists' mobility may be attributable to various factors including cultural influences, economic opportunities, and research collaboration networks, yet all of which warrant further investigation and validation.

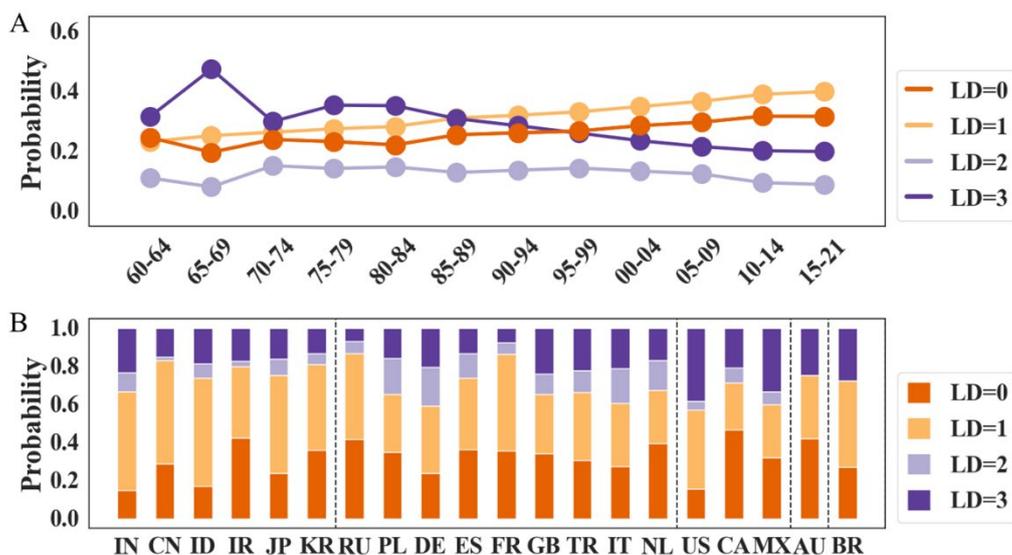

**Figure 3. The probability of mobility across level distance at different levels. A.** Average probability of mobility of different cohorts categorized by different distance levels. **B.** Probability of mobility at different distance levels based on home countries during 2015 and 2021. According to our SMART model, a scientist's decision to move is influenced by their home country (the country in which they resided before relocating). The willingness to cross different levels of borders varies among different countries. For instance, in France, scientists are more inclined to relocate within the country by choosing universities in different cities.

**Level distance implies geographical distance**
In human mobility, location proximity (i.e., geographical distance) between the origin and destination often constitutes a significant determinant of scientists' decisions to migrate [38, 39]. In our model, the probability of mobility at a certain level distance is employed as a measure of the likelihood of scientists' relocation to other locations, given their home countries. This measure inherently accounts for geographical proximity and incorporates additional factors influencing movement. These include cultural differences, salary disparities, travel costs, and potential political considerations, many of which are closely linked with distance.

Our analysis, based on the defined level distance, shows that mobility distances at various levels align sequentially with the levels' magnitude (see Fig. 4A). Thus, the parameter in our model encapsulates geographic distance information. Furthermore, we compared the probabilities of different levels of mobility occurring within countries between 2015 and 2021, as the function of the average mobility distances at these levels. Intriguingly, Fig. 4B indicates that at level distances 0 and 2, the mobility distance and the likelihood of this mobility exhibit a negative correlation, which suggests that scientists tend to move to locations with shorter geographical distances when conducting movements between institutions in the same city (level distance is 0) or between institutions in different countries but on the same continent (level distance is 2). This can be explained as follows. Scientists often target new positions within their current city, favoring closer institutions that are more familiar or with which they might have had previous collaborations, thereby increasing the chances of securing a position there. Regarding international mobility, our dataset focused on 200 major cities, representing only a limited number of countries. The observed frequent intra-continental movement tend to occur within culturally similar regions, leading to the noted negative correlation between geographical distance and mobility level.

The positive correlation between the probability of mobility at level distance 3 and the geographical distance observed can be attributed to the movements primarily conducted between regions such as Asia-Europe, Asia-America, and Europe-America, whereas they are less common between regions such as South America-North America and Europe-Africa. This increased frequency of intercontinental movements in the former regions can be attributed to greater average distances compared to the latter regions, leading to a higher likelihood of occurrence as distance increases.

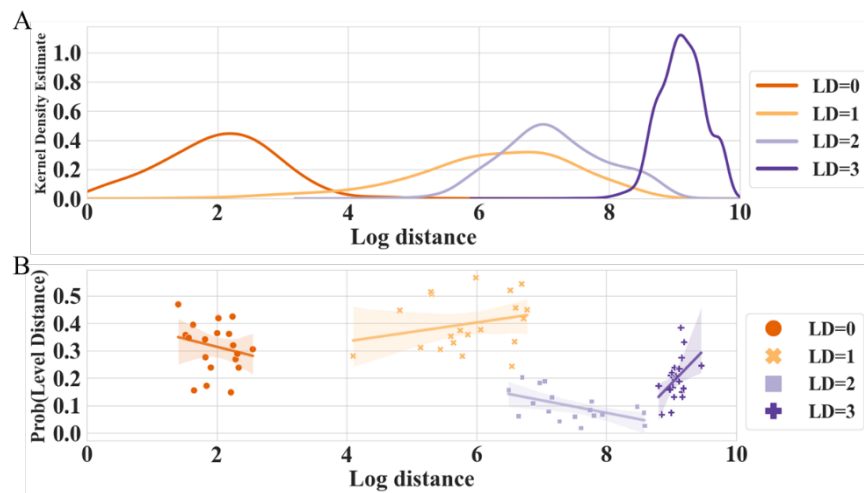

**Figure 4. The distance of scientists' mobility and its relationship to probability of mobility at different distance levels. A.** The kernel distribution estimates of distance of scientists' mobility at different levels of region. **B.** The correlation between geographical distance and probability of mobility at different distance levels given home countries during 2015 and 2021. Color indicates mobility at different levels or

with different level distance. Lines show the linear regression trends with 95% confidence intervals.

**Attractiveness reveals the scientific status of regions**

What traits make different regions appealing at each mobility level? Although scientists' mobility constitutes a choice driven by a comprehensive assessment of factors conducive to their personal development, we explored the potential collective characteristics of the regions that can be measured from our data. Specifically, we examine the (i) impact of the number of institutions within a region which is the number of opportunities the region can provide to the mover, (ii) the number of researchers who are potential collaboration candidates, and (iii) the number of publications which is the scientific capital within the region at a specific level.

The results in Fig. 5(A, C, E) reveal a positive correlation between the attractiveness of countries on different continents and the number of institutions, number of authors, and number of publications within the region. Fig. 5(B, D, F) represents cases at the city level. The three indicators within each city across national levels exhibit a positive correlation with attractiveness. Therefore, regions with higher relative research proficiency and larger scale have a greater ability to attract external scholars.

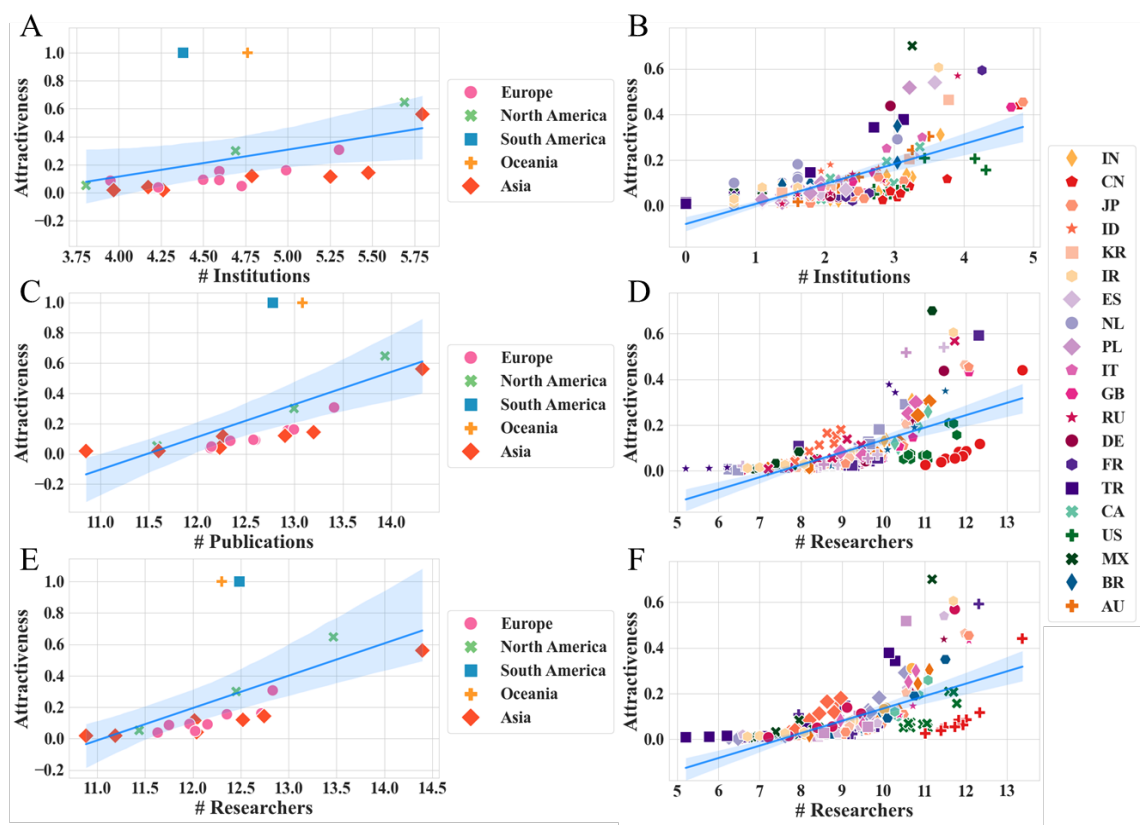

**Figure 5. The correlation between region attractiveness and scientific status in different levels in the last cohort (11th: 2015–2021). A, C, E** display the relationship between attractiveness and number of institutions, publications, and researchers,

respectively, at the continent level. **B, D, F** display the relationship between attractiveness and number of institutions, publications, and researchers, respectively, at the country level. Marker color and shape distinguish among regions. Lines show the linear regression trends with 95% confidence intervals.

## Discussion

Studies have quantitatively demonstrated the presence of scales in human mobility behavior through empirical investigations [13, 40]. We found the similar pattern in scientific mobility, as Fig. 1(C, D, and E) shows, scientists' internal mobility within regions, such as cities, nations, or continents, is greater than their external movements to other regions. This study employed the SMART model to explore the intrinsic patterns of scientists' mobility. Employing maximum likelihood estimation, we fit the obtained trajectories of 2.03 million scientists' career transitions over the period 1960–2021, segmented into 5-year intervals. The resulting parameters signify the significance of considering administrative units as the fundamental entity for studying scientist mobility. For instance, the attractiveness parameter reveals the relative appeal of a region compared to regions within the same parent region. The definition of level distance determines its probability of occurrence in relation to a scientist's home country. Level distance is correlated with geographical distance. However, as shown in Fig. 4, we found that the correlation between the geographical distance and the mobility likelihood is varies across the different level distances, with the negative correlations found in level distance 0 and 2, which is not discovered before.

The concept of "level distance" in the context of international scientists' mobility encompasses various factors, such as culture differences, language disparities, and economic variations. These elements play a crucial role in shaping the ease of a scientist's transition to a new academic or research environment. The extent of cultural differences, linguistic challenges, and economic disparities between the scientist's home country and the host institution can significantly influence the decision to relocate. Meanwhile, "attractiveness" in the context of our model is associated with the prestige of the host institution and the potential for collaboration opportunities. A highly esteemed institution with a track record of research excellence and a robust collaborative network may be more attractive to scientists seeking new positions. The allure of an institution is not based only on its reputation, but also on the potential for meaningful collaborations and research synergies, which can further enhance a scientist's career and impact within their field.

The results of this study are based on maximum likelihood estimation to derive parameters for the defined trajectory transition probability function. The values of each parameter correspond to the observed frequency within the defined parameter context over the observation period. Thus, we did not perform artificial trajectory fitting for the model, because the fit aligned perfectly with the observed phenomena. A limitation of this study is that we selected only the most active 200 cities globally in scientific endeavors, that is, the top 200 cities in terms of publication volume, potentially omitting emerging active scientific cities. Additionally, our adoption of a 5-year interval for trajectory segmentation reduces the computational complexity of the parameter estimation albeit at the expense of accuracy. This is because of the relatively short trajectories of scientists' movements, which may span several years or even an entire career without institutional changes. Future endeavors could involve larger-scale computational methods or finer geographical divisions to further explore the model proposed in this study.

In this study, we introduced an innovative approach that employs actual administrative regions like cities and countries as units to analyze scientists' mobility patterns. This method moves beyond traditional models that primarily focus on physical distance [18, 19] or those that are complex and computationally demanding [21, 22]. We discovered that scientists' relocation decisions vary based on the administrative area. For instance, within the same city, proximity is often preferred, but when considering moves to another continent, distance becomes less critical. We also creatively applied the concept of 'attractiveness' to evaluate the appeal of different areas to scientists at the same administrative level, revealing significant variations in attractiveness. This suggests that scientists' choices are influenced not just by proximity, but also by the relative appeal of different regions.

Since the 1960s, Asia has experienced notable changes in attracting talent and scientific mobility. This shift is largely due to rapid economic growth, enhanced investments in education and research, and diverse immigration policies. Countries like China and South Korea have emerged as more attractive destinations for talents, owing to their improved higher education systems and research infrastructures, as shown in the estimated attractiveness coefficients in Fig. 2. Enhanced regional collaboration, especially through ASEAN, has furthered mobility and integration within Asia. Nevertheless, challenges like brain drain, competition among Asian nations, and issues related to political and academic freedom persist. Despite these challenges, Asia's role in the global scientific community has been bolstered by technological advancements and innovation, though it continues to navigate both internal and international challenges and competition.

By considering both geographical distance and area attractiveness, our model provides a more comprehensive understanding of scientists' mobility patterns, showing that their decisions are influenced by a combination of factors. This insight is crucial for institutions and policymakers focused on attracting and retaining scientific talent, offering new perspectives that challenge traditional assumptions and open up fresh avenues for research in scientific mobility. The implications of our findings extend to practical concerns such as HR policies in universities and governmental policy incentives. By decoding mobility patterns, universities can better strategize to attract and retain scientific talent, considering factors like institutional allure and collaborative opportunities. They may need to evolve research conditions and provide competitive offerings reflective of regional economic conditions. Government policies aiming to ease mobility could address cultural and linguistic barriers and offer economic incentives to draw international talent, enhancing scientific vibrancy.

This emphasis on internal development can also inform national policies to bolster resources in up-and-coming scientific cities to curb brain drain. The knowledge of such internal mobility patterns can aid in creating a more equitable scientific community across different regions. Hence, the study's insights are pivotal for developing HR strategies and governmental policies, which could potentially lead to an optimally functioning and equitable global scientific ecosystem.

# Methods

**Data**

Through data gathering and processing, we obtained the career trajectories of scientists from 1960 to 2021, covering 75,560 academic institutions. We selected 8,366 institutions, which collectively published 95% of papers. By gathering and processing data from OpenAlex, we filtered out all publication records, including information on authors, host institutions, and publication dates, between 1960 and 2022. After excluding organizations that lacked complete geographic information, the dataset contained 7629 institutions.

We narrowed our focus further by selecting only cities with the top 10 highest represented academic institutions in countries with the top 20 highest numbers of academic institutions in our dataset. Finally, there are four levels of the geographical hierarchical model: 2589 institutions are in level 0, 200 cities are in level 1, 20 countries are in level 2, and five continents are in level 3 (Fig. 1A).

To investigate the mobility patterns of scientists within these institutions over time periods starting from 1960 to 2021, we divided the data into lengths of 5 years. We defined mobility as the shift in the publications associated with a scientist's host institution in each timeframe. We constructed a link table or adjacent matrix for each 5-year timeframe, with rows representing the origin institution and columns representing the destination institution; each cell contained the number of people who moved from one institution to another during that timeframe. Note that 1960–1965 represents cohort 0, while 2015–2021 represents cohort 11.

In the main paper, we present the estimation results of the container model applied to the mobility data used in our study, which encompass the movements of scientists, where the origin corresponds to any institution within the top 20 countries, and the destination is an institution located in one of the top 10 cities within the top 20 countries. In this context, the term "top" refers to countries that possess the highest number of institutions.

**Maximum likelihood estimation of $p(d,c)$**

Note that for a certain country $c$, $p(d,c)$ is independent of all movements that do not start from $c$. Thus, we need only show how to estimate $p(d,c)$ for one country, and the method can also be applied to other countries.

Under our settings, the level distance $d$ can only take the values of 0,1,2,3. For a fixed country $c$, write $p_c = \bigl(p(0,c), p(1,c), p(2,c), p(3,c)\bigr)^T \in \mathbb{R}^4$; then, the likelihood is

$$L(p_c) = p(0,c)^{n_0} p(1,c)^{n_1} p(2,c)^{n_2} p(3,c)^{n_3} \qquad (2)$$

and the log-likelihood function is

$$l(p_c) = n_0 \ln p(0,c) + n_1 \ln p(1,c) + n_2 \ln p(2,c) + n_3 \ln p(3,c) \quad (3)$$

where $n_i$ denotes the number of movements with level distance $i$ starting from country $c$.

The objective for MLE is

$$\max l(p_c)$$
$$\text{s.t.} \, 0 < p(i,c) < 1, i = 0,1,2,3; \sum_{i=0}^{3} p(i,c) = 1. \quad (4)$$

By the Kuhn–Tucker (KT) conditions, we can obtain the optimal solution. The Lagrangian function is

$$L(p_c, \lambda, v) = l(p_c) - \sum_{i=0}^{3} \lambda_i p(i,c) + \sum_{i=0}^{3} u_i (p(i,c) - 1) + v(\sum_{i=0}^{3} p(i,c) - 1). \quad (5)$$

Taking the partial derivative of $p_c$ and setting it to zero, we obtain

$$\frac{n_i}{p(i,c)} - \lambda_i + u_i + v = 0, i = 0,1,2,3. \quad (6)$$

Because $p(i,c)$ can be neither 0 nor 1, the KT conditions force all $\lambda_i$ and $u_i$ to be 0. Then we have

$$\frac{n_0}{p(0,c)} = \frac{n_1}{p(1,c)} = \frac{n_2}{p(2,c)} = \frac{n_3}{p(3,c)} = -v. \quad (7)$$

We also have $\sum_{i=1}^{3} p(i,c) = 1$, and we can solve the optimal solution:

$$\hat{p}(i,c) = \frac{n_i}{\sum_{j=0}^{3} n_j}. \quad (8)$$

**Maximum likelihood estimation of attractiveness**

For all containers in a parent container, $k_{l+1}$, the likelihood function and log-likelihood function are:

$$L(a) = \prod_{i \in k_{l+1}} a(i)^{m_i}, \quad (9)$$

$$l(a) = \sum_{i \in k_{l+1}} m_i \ln a(i), \quad (10)$$

where $m_i$ is the number of movements to the destination in container $i$. The objective of MLE is

$$\max l(p_c)$$
$$\text{s.t.} \, 0 < a(i) < 1, i \in k_{l+1}; \sum_{i \in k_{l+1}} a(i) = 1. \quad (11)$$

The Lagrangian function is

$$L(a, \lambda, v) = l(a) - \sum_{i \in k_{l+1}} \lambda_i a(i) + \sum_{i \in k_{l+1}} u_i (a(i) - 1) + v(\sum_{i \in k_{l+1}} a(i) - 1). \quad (12)$$

Taking the partial derivative of $a$ and setting it to zero, we have

$$\frac{m_i}{a(i)} - \lambda_i + u_i + v = 0, \forall i \subset k_{l+1} \quad (13)$$

Similarly, the KT conditions forces $\lambda_i$ and $u_i$ to be 0. Then, we have:

$$\frac{m_i}{a(i)} = -v, \forall i \subset k_{l+1} \tag{14}$$

Because $\sum_{i \in k_{l+1}} a(i) = 1$, the optimal solution is

$$\widehat{a}(i) = \frac{m_i}{\sum\limits_{j \in k_{l+1}} m_j} \tag{15}$$

**The meaning of the level distance and attractiveness**
The SMART model employs a five-year timeframe to illustrate the movement patterns of scientists, focusing on changes in their affiliated institutions as reflected by their publication history. For instance, if a scientist published an article affiliated with Peking University in China in 1968, followed by another article affiliated with Northwestern University in the USA, the scientist effectively transitioned from Peking University in Beijing, China (Asia) to Northwestern University in Evanston, USA (North America). The pre-move location is defined as (Peking University, Beijing, China, Asia), while the post-move location is (Northwestern University, Evanston, USA, North America).

In determining scientists' decisions to shift between institutions, we assume their willingness to move across city, country, or continental borders is rooted in their "home country," i.e., the country they were in before the move. Consequently, we employ the concept of "level distance" to quantify the effort involved when scientists undergo such mobility. This "level" refers to the initial administrative level difference between the scientist's pre-move and post-move locations of institutions. In the earlier example, the level distance would be 3 due to the continents' disparity between Peking University and Northwestern University. Conversely, if a scientist moves from Peking University to Tsinghua University, the level distance would be 0, given both universities are situated in Beijing.

The likelihood of a scientist traveling a distance $d(j,k)$ starting from country $c$ is denoted by $p(d(j,k), c)$. This implies that when a scientist is in university j, located in country c, the probability of their mobility to another university $k$ in the same city $(d(j,k) = 0)$, country $(d(j,k) = 1)$, continent $(d(j,k) = 2)$, or a different continent $(d(j,k) = 3)$ as university $j$. Once the level distance of a mobility is determined, the scientists' choice of destination follows a hierarchical pattern. For instance, a scientist from Peking University deciding to move to the USA signifies the initial level distance choice. The subsequent choice is to opt for continents other than Asia, then specifically selecting North America over other continents, followed by choosing the USA within North America and subsequently selecting Evanston in the USA and Northwestern University within Evanston.

The coefficient describing a scientist's selection of a sub-region under a parent region is termed as "attractiveness." This measure signifies the probability of choosing that region among all regions nested within the same parent region. For instance, between 1995 and 1999, the attractiveness values for the USA, Canada, and Mexico were 0.63, 0.32, and 0.05, respectively. This indicates that when scientists decide to move to North America from a location in other continents, the probability of choosing the USA as their destination is 0.63.

**The null model**
We calculated the simulated number of movements across different institutions using a null model in which all the movements were randomized. Let $S_1$ denote the set of institutions within the top 30 countries, and $S_2$ denote the set of institutions located in one of the top 10 cities in the top 20 countries. The simulated number of movements from institution $a \in S_1$ to institution $b \in S_2$ is denoted as $m_{a,b}'$ and calculated by

$$m_{a,b}' = \frac{D_a^{out} D_b^{in}}{\sum_{x \in S_1} D_x^{out}}, \tag{16}$$

where $D_a^{out}$ denotes the number of movements starting from $a$ with the destination being any institution in $S_2$, and $D_b^{in}$ denotes the number of movements starting from any institution in $S_1$ with the destination being .

**Acknowledgments**
Y.M. was supported by the National Natural Science Foundation of China (No. NSFC62006109 and NSFC12031005), and partially supported by the Support Plan Program of Shenzhen Natural Science Fund No. 20220814165010001. R.S. acknowledges support by Villum Fonden through the Villum Young Investigator program (project number: 00037394). The computation in this study was supported by the Center for Computational Science and Engineering of SUSTech.

**Competing Interests:**
The authors declare no competing interests in this work.

**Data and Code Availability**
The OpenAlex data is publicly available [34]. The code and data for reproducing our main results will be shared in a permanent repository.

# Supplementary Information for
# Quantifying the hierarchical scales of scientific mobility

Yurui Huang[1†], Langtian Ma[1†], Chaolin Tian[1], Xunyi Jiang[1], Roberta Sinatra[2,3,4], Yifang Ma[1*]

[1]Department of Statistics and Data Science, Southern University of Science and Technology, Shenzhen 518055, Guangdong, China

[2]Copenhagen Center for Social Data Science (SODAS), University of Copenhagen, Denmark

[3]Networks, Data, and Society (NERDS) group, IT University of Copenhagen, Denmark

[4]ISI Foundation, Turin, Italy

*Email: mayf@sustech.edu.cn

†These authors contributed equally to this work.

# Supplementary Figures and Tables

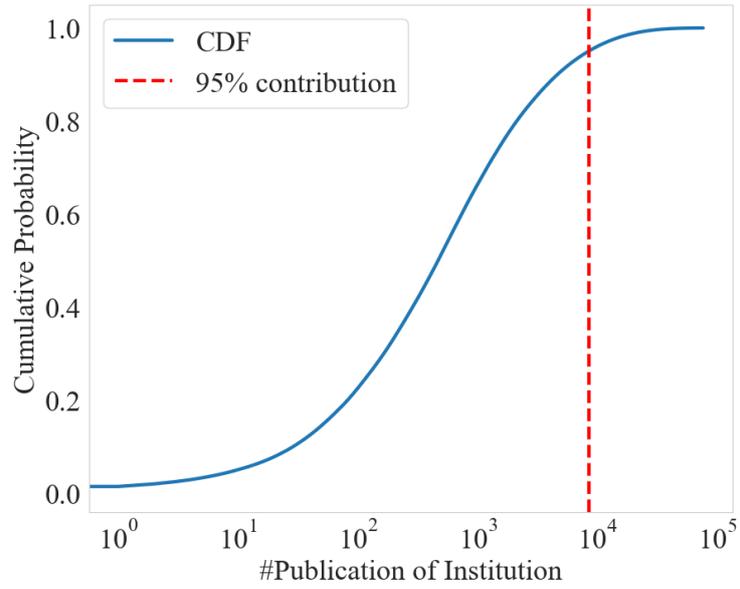

**Figure S1.** Cumulative Distribution Function (CDF) of number of publications for scientific institutions. 95% publications are contributed by 8,366 institutions.

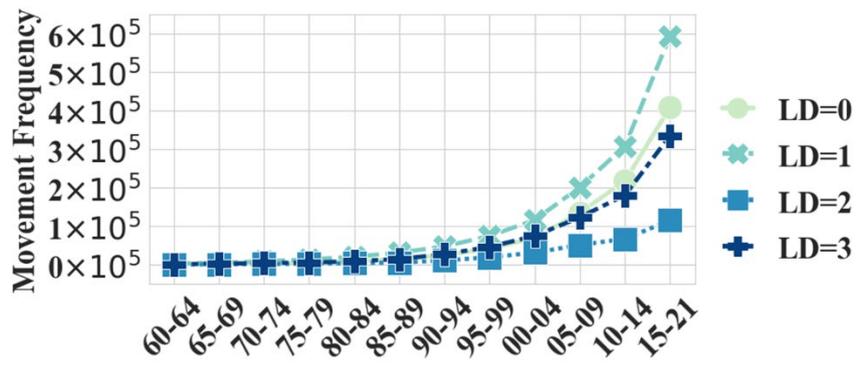

**Figure S2.** Distribution of mobility frequency in different level.

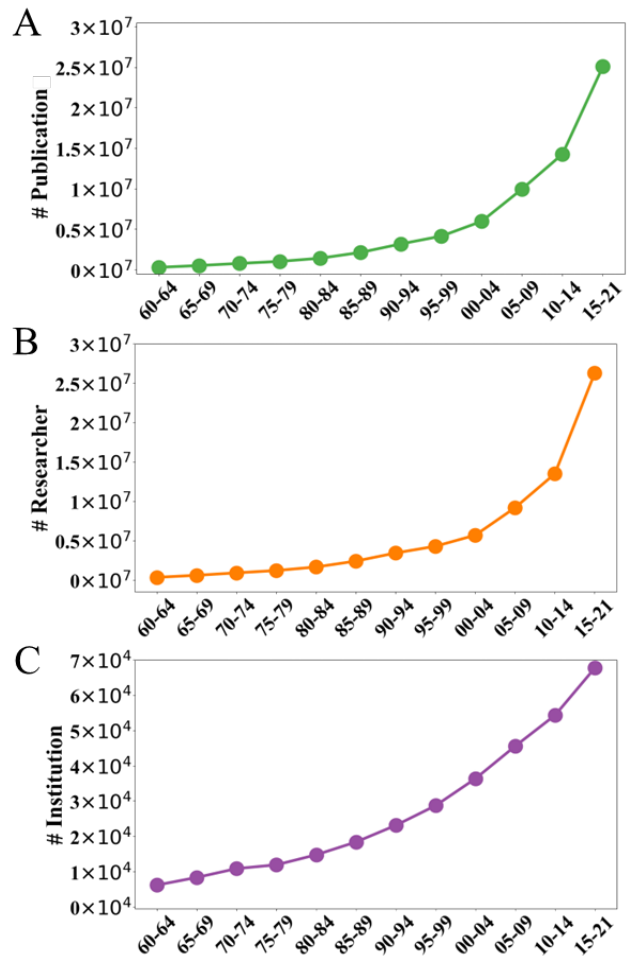

**Figure S3.** Changing Trends in numbers of publications, researchers, and institutions from 1960 to 2021

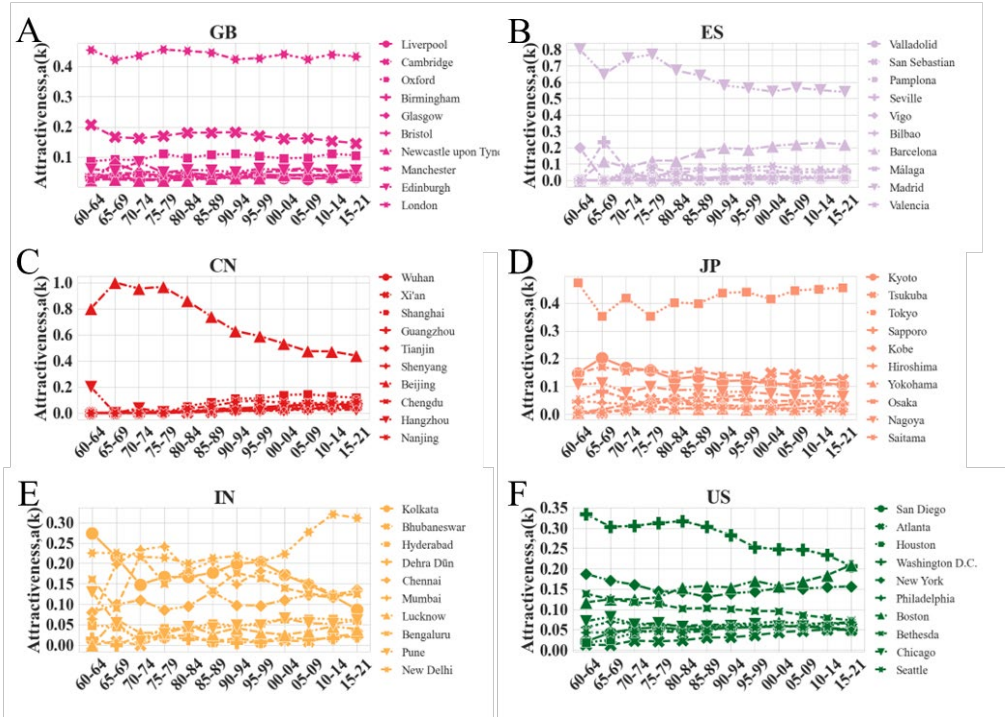

**Figure S4.** The attractiveness coefficients in country levels of Region Model. The attractiveness coef. are presented in country level from cohort 0 to cohort 12 of Great Britan(A), Spain(B), China(C), Japan(D), India(E), and the United States(F).

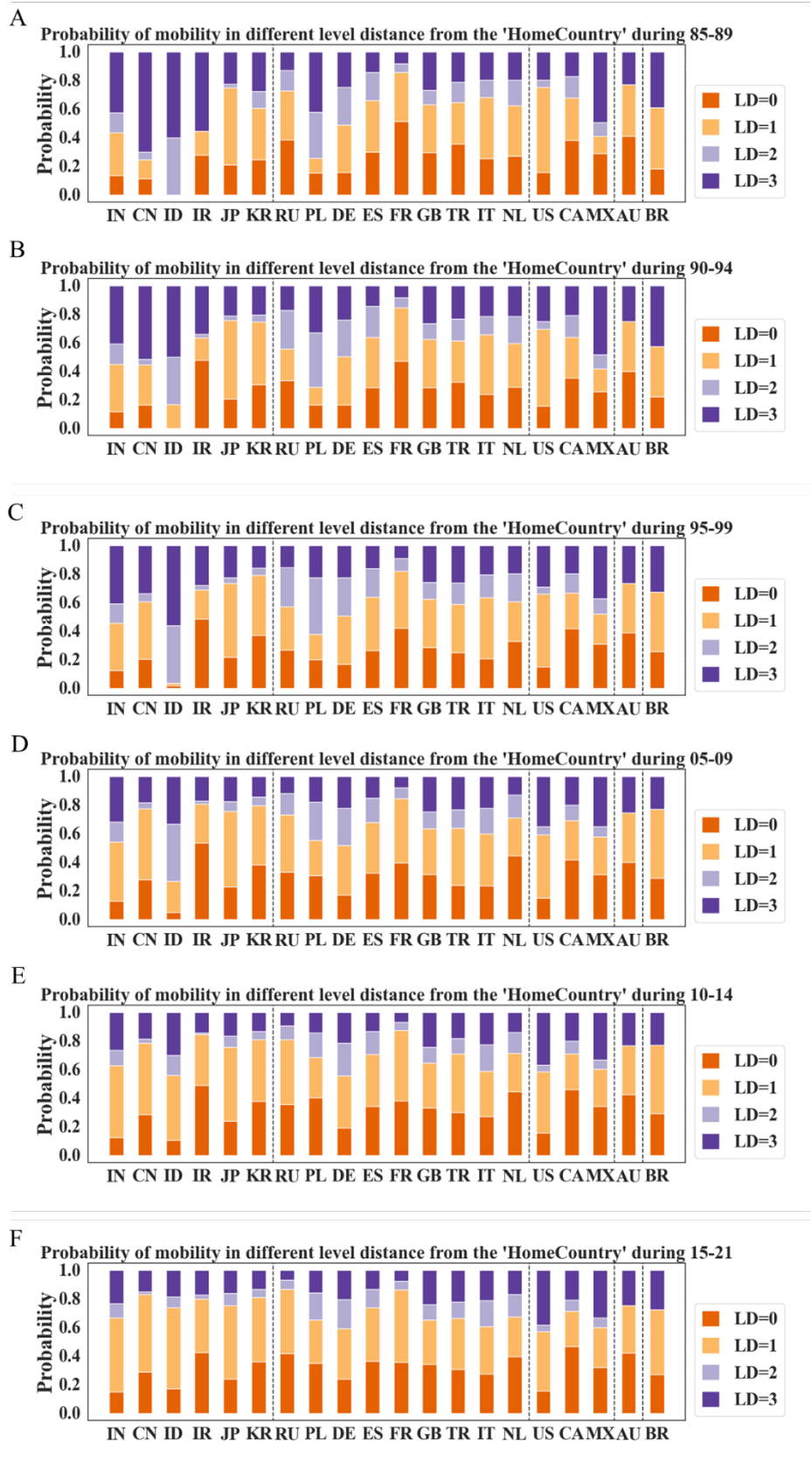

**Figure S5. The probability of level distance coefficients in four levels of Region Model in different cohort.** 1985-1989(A), 1990-1994(B), 1995-1999(C), 2005-2009(D), 2010-2014(E), 2015-2021(F).